\documentclass[]{article}

\usepackage{hyperref}
\usepackage[usenames,dvipsnames,svgnames,table]{xcolor}
\usepackage{authblk}

\usepackage{mathtools}
\usepackage{todonotes}
\usepackage{listings}
\usepackage[normalem]{ulem}

\usepackage{acronym}
\acrodef{mas}[MAS]{Multi-Agent System}
\acrodef{map}[MAP]{Multi-Agent Planning}
\acrodef{maps}[MAP]{Multi-Agent Plan}
\acrodef{mapddl}[MA-PDDL]{Multi-Agent-Planning Domain Definition Language}
\acrodef{ap}[AP]{Aggregate Programming}
\acrodef{fc}[FC]{Field Calculus}
\acrodef{dsl}[DSL]{Domain-Specific Language}
\acrodef{iiot}[IIoT]{Industrial Internet of Things}
\acrodef{bda}[BDA]{Big Data Analytics}
\acrodef{dml}[DML]{Data Management Layer}
\acrodef{ue}[UE]{User Equipment}

\lstdefinelanguage{fcpp}{
  keywords={class,struct,template,public,return,if,for,while},
  keywordstyle=\color{blue}\textbf,
  otherkeywords={=>,<-,<\%,<:,>:,\#,@},
  keywordstyle=[2]\color{red},
  keywords=[2]{old,nbr,oldnbr,spawn},
  keywordstyle=[3]\color{violet},
  keywords=[3]{abf,min_hood,fold_hood,nbr_dist,and,or},
  keywordstyle=[4]\color{Emerald},
  keywords=[4]{double,bool,int,P,F,T,U,G,Ts,node_t,trace_t,device_t,real_t,trace_call,field,tuple,array,vec},
  keywordstyle=[5]\color{Brown},
  keywords=[5]{INF,FUN,GEN,ARGS,CODE,CALL,true,false},
  sensitive=true,
  morecomment=[l]{//},
  morecomment=[n]{/*}{*/},
  commentstyle=\color{darkgreen},
  morestring=[b]",
  morestring=[b]',
  morestring=[b]"""
}
\newcommand{\clsys}{\mathit{N}_\mathit{C}}
\newcommand{\mobn}{\mathit{N}_\mathit{M}}
\newcommand{\fixn}{\mathit{N}_\mathit{F}}
\newcommand{\gwyn}{\mathit{N}_\mathit{G}}
\newcommand{\edgenet}{\mathcal{N}_\mathit{EDGE}}
\newcommand{\intnet}{\mathcal{N}_\mathit{INET}}
\newcommand{\dists}{\mathit{S}_\mathit{DIST}}
\newcommand{\params}{\mathit{S}_\mathit{PARAM}}
\newcommand{\nssafe}{\mathit{N}_\mathit{SAFE}}
\newcommand{\nscons}{\mathit{N}_\mathit{CONS}}
\newcommand{\nsprod}{\mathit{N}_\mathit{PROD}}

\newcommand{\prodnode}{\mathit{n_{PROD}}}
\newcommand{\consnode}{\mathit{n_{CONS}}}

\definecolor{mylightgray}{rgb}{0.97,0.97,0.97}
\definecolor{mygreen}{rgb}{0,0.6,0}
\definecolor{mygray}{rgb}{0.5,0.5,0.5}
\definecolor{mymauve}{rgb}{0.58,0,0.82}

\lstdefinelanguage{hfc}{
	basicstyle=\ttfamily, 
	frame=single,
	basewidth=0.5em,
	sensitive=true,
	morestring=[b]",
	morecomment=[l]{//},
	morecomment=[n]{/*}{*/},
	commentstyle=\color{OliveGreen},
	keywordstyle=\color{blue}\textbf, keywords={def}, otherkeywords={=>,||,\&\&,!,<=,==},
	keywordstyle=[2]\color{red}\textbf, keywords=[2]{rep,nbr,if,share},
	keywordstyle=[3]\color{violet}, keywords=[3]{mux,dist,lag,sumHood,countHood,allHood,anyHood,locHood,minHood,maxHood,foldHood,fst,snd},
	keywordstyle=[4]\color{orange}\textbf, keywords=[4]{spawn},
	keywordstyle=[5]\color{blue}, keywords=[5]{false,true,infinity,null}
}

\lstdefinelanguage{cpp}{
	basicstyle=\ttfamily\scriptsize,
	keywords={typename,auto,using,namespace,include,class,struct,template,public,return,if,for,while},
	keywordstyle=\color{blue}\textbf,
	otherkeywords={=>,<-,<\%,<:,>:,\#,@},
	keywordstyle=[2]\color{Emerald},
	keywords=[2]{double,bool,int,T,Ts},
	keywordstyle=[3]\color{Brown},
	keywords=[3]{true,false},
	keywordstyle=[4]\color{violet},
	keywords=[4]{MAIN},
	sensitive=true,
	morecomment=[l]{//},
	morecomment=[n]{/*}{*/},
	commentstyle=\color{OliveGreen},
	morestring=[b]",
	morestring=[b]',
	morestring=[b]"""
}

\newcommand{\BNFcce}{{\bf ::=}}
\newcommand{\BNFmid}{\;\bigr\rvert\;}

\newcommand{\FUNCTION}{\mathtt{F}}
\newcommand{\type}{t}
\newcommand{\btype}{bt}
\newcommand{\ttype}{tt}
\newcommand{\tvar}{T}
\newcommand{\ap}[1]{\texttt{<}#1\texttt{>}}
\newcommand{\ftype}[1]{\mathtt{field}\ap{#1}}
\newcommand{\defK}{\mathtt{FUN}}
\newcommand{\argK}{\mathtt{ARGS}}
\newcommand{\codeK}{\mathtt{CODE}}
\newcommand{\callK}{\mathtt{CALL}}
\newcommand{\retK}{\mathtt{return}}
\newcommand{\toK}{\texttt{->}}
\newcommand{\nodeK}{\mathtt{node}}

\newcommand{\e}{\mathtt{e}}
\newcommand{\bname}{\mathtt{b}}
\newcommand{\cname}{\mathtt{c}}
\newcommand{\dname}{\mathtt{d}}
\newcommand{\fname}{\mathtt{f}}
\newcommand{\pname}{\mathtt{p}}
\newcommand{\oname}{\mathtt{o}}
\newcommand{\uname}{\mathtt{u}}
\newcommand{\xname}{\mathtt{x}}

\newcommand{\lvalue}{\ell}
\newcommand{\oldK}{\mathtt{old}}
\newcommand{\nbrK}{\mathtt{nbr}}

\newcommand{\spawnK}{\mathtt{spawn}}
\newcommand{\truevalue}{\mathtt{true}}
\newcommand{\falsevalue}{\mathtt{false}}

\begin{document}


  \title{Aggregate Processes as Distributed Adaptive Services \\
    for the Industrial Internet of Things}

%
%
%

 \author[1,2]{Lorenzo Testa}

 \author[1]{Giorgio Audrito}

  \author[1]{Ferruccio Damiani}

  \author[1]{Gianluca Torta}

  \affil[1]{Dipartimento di Informatica,  University of Turin, Turin, Italy}
  \affil[2]{Concept Reply, Turin, Italy}

 \maketitle
 
\begin{abstract}
  The Industrial Internet of Things (IIoT) promises to bring many benefits, including increased productivity, reduced costs, and increased safety to new generation manufacturing plants.
  The main ingredients of IIoT are the connected, communicating devices directly located in the workshop floor (far edge devices), as well as edge gateways that connect such devices to the Internet and, in particular, to cloud servers.
  The field of Edge Computing advocates that keeping computations as close as possible to the sources of data can be an effective means of reducing latency, preserving privacy, and improve the overall efficiency of the system, although building systems where (far) edge and cloud nodes cooperate is quite challenging.
  In the present work we propose the adoption of the Aggregate Programming (AP)  paradigm (and, in particular, the ``aggregate process'' construct) as a way to simplify building distributed, intelligent services at the far edge of an IIoT architecture.
  We demonstrate the feasibility and efficacy of the approach with simulated experiments on  FCPP (a C++ library for AP), and with some basic experiments on physical IIoT boards running an ad-hoc porting of FCPP.
\end{abstract}




\section{Introduction}
The \ac{iiot}, namely the concepts and technologies of IoT applied to (smart) industry, is gaining increasing interest as it promises to improve productivity and safety in the workplace through the collection, analysis and exploitation of large amounts of data from the workshop floor. While there is no single definition of \ac{iiot}, the following one (taken from \cite{khan:CEE20}) can be a useful reference for our purposes: ``\ac{iiot} is the network of intelligent and highly connected industrial components that are deployed to achieve high production rate with reduced operational costs through real-time monitoring, efficient management and controlling of industrial processes, assets and operational time.''

\begin{figure}[tb]
	\centering
	\includegraphics[width=\linewidth]{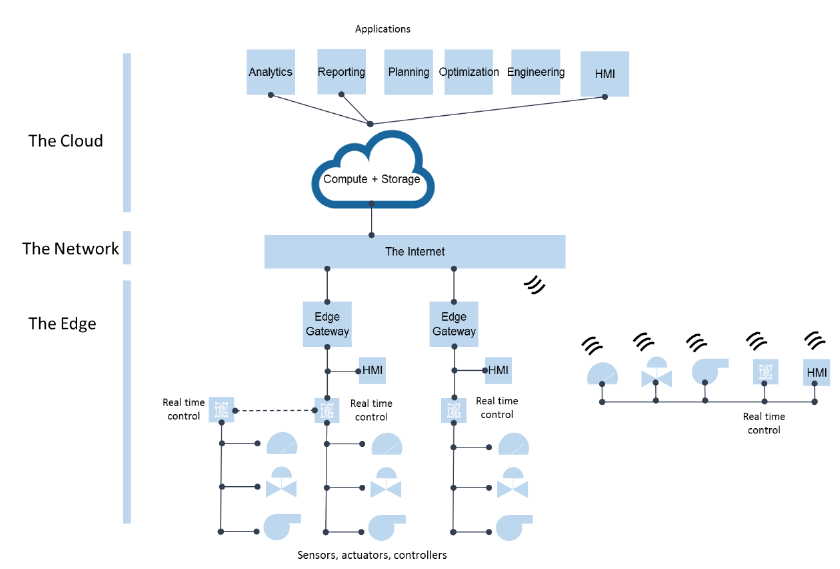}
	\vspace{-10pt}
	\caption{Reference Architecture of the \ac{iiot} (from \url{https://en.wikipedia.org/wiki/Industrial_internet_of_things}).} \label{fig:iiot-arch}
\end{figure}

From this definition we can deduce that connectivity and data collection are the main enabling elements of the \ac{iiot}. Furthermore, the reference architectures of the \ac{iiot} also stress the need of accessing high performance computational and storage nodes at a higher level, usually in the cloud. Again, we find several slightly different definitions of the \ac{iiot} architecture (see, e.g., \cite{khan:CEE20, sisinni:TII18}); all of which, however, share most aspects with one depicted in Figure \ref{fig:iiot-arch}.

\begin{figure}[tb]
	\centering
	\includegraphics[width=0.75\linewidth]{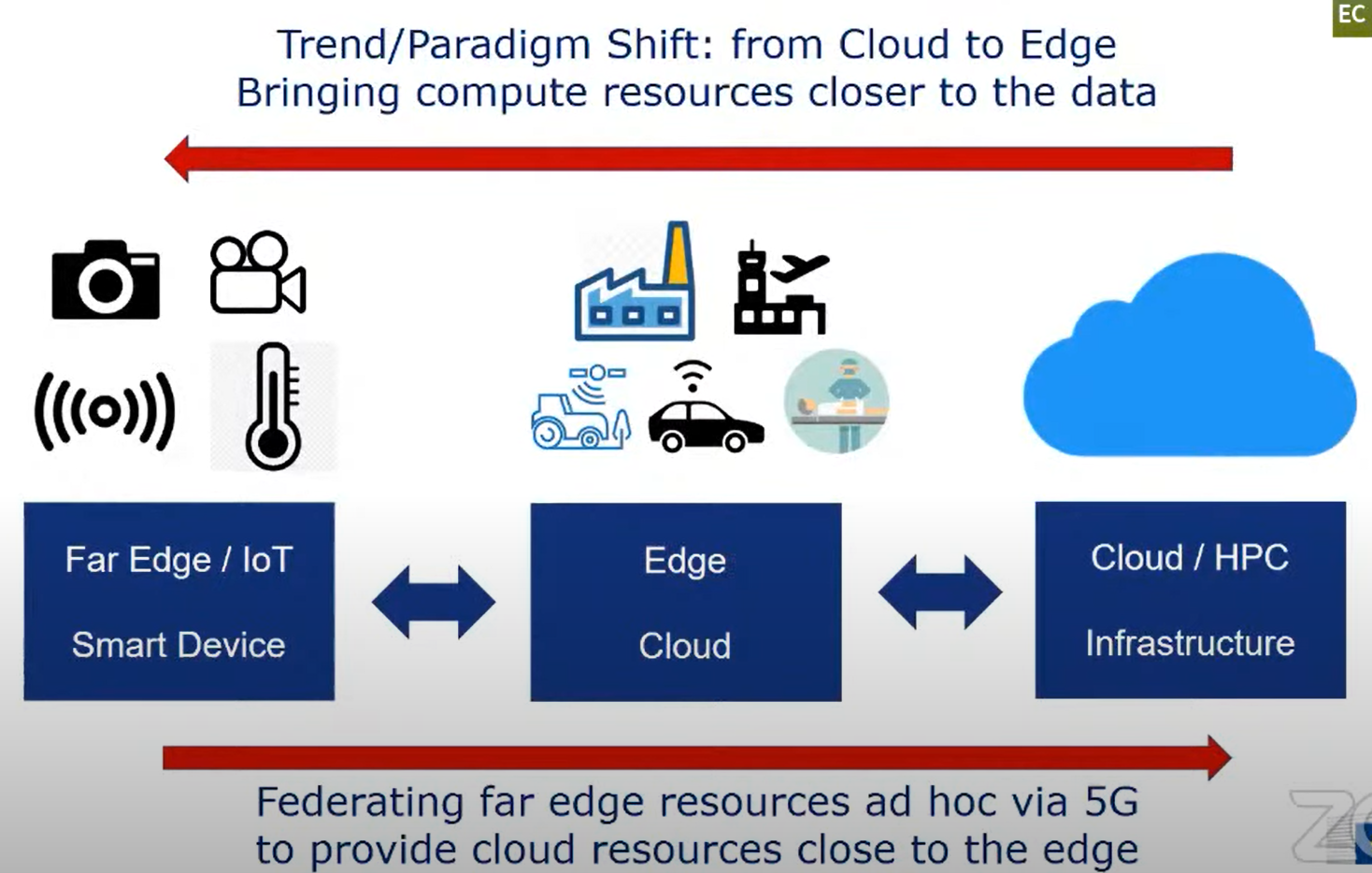}
	\vspace{-10pt}
	\caption{Cloud-Edge-IoT Orchestration.}
        \label{fig:iot-ec}
\end{figure}

We further partition the bottom layer in the figure into two sublayers: the {\em Far Edge} layer which contains the connected devices in the workshop floor, including: machines, sensors, actuators, real-time controllers, Human-Machine-Interface (HMI) units; and the {\em Edge gateways} that act as bridges towards the outside world. Then, the information collected by the gateways crosses the {\em Network} layer to reach the {\em Cloud} layer, where it is processed by rich applications for analysis, planning, optimization, etc. Of course the flow can also be reversed, so that plans and decisions created at the {\em Cloud} layer can reach the {\em Far Edge}.

A well known potential problem with this scheme is that pushing all the high-level functions to the cloud (i.e., far from the workshop floor), can have a negative impact on the reliability, cost, scalability and latency of the system. The field of {\em edge computing} aims at addressing the problem by moving some computations closer to the edge, notably moving (part of) the data processing closer to where the data is produced (see Figure \ref{fig:iot-ec}, image from the presentation of
the Horizon Europe Destination 3:
	Call ‘World leading data and computing technologies 2022’.).\footnote{The presentation
	(given by Jan Comar at the Horizon Europe Cluster 4 - Digital, Industry \& Space Info Day 2021 \url{https://ec.europa.eu/info/research-and-innovation/events/upcoming-events/horizon-europe-info-days/cluster-4_en}) can be found at \url{https://www.youtube.com/watch?v=SevWyhwaEwE}.}  The term {\em edge computing} does not prescribe exactly how close to the {\em things} (i.e., devices) the computation should be moved, and how much powerful should be the edge computing nodes. Thus, referring to Figure \ref{fig:iiot-arch}, the term can be applied to all the following (and quite diverse) situations: computations running at servers in the {\em Network} layer that are only a few hops from the {\em Edge Gateway}s; computations running in the {\em Edge Gateway}s themselves; and computations running in the {\em Far Edge} layer, exploiting the constrained memory and computing capabilities of smart devices.

In this paper we focus on the latter meaning of edge computation, whereby offloading some of the computations to the {\em Far Edge} layer can make them more robust, adaptive and responsive. In particular, we propose an approach that makes it possible to build distributed services for the \ac{iiot} running on highly resource-constrained devices.  At the core of our approach lies the possibility of running distributed computations -- defined as processes in the \acf{ap} paradigm \cite{bpv:aggregate:programming} and in the foundation of this paradigm provided by the \acf{fc}~\cite{avdpb:TOCLhfc} -- on IIoT devices by relying on the FCCP library, a C++ implementation of \ac{fc} \cite{a:fcpp}. The \ac{ap} paradigm is suitable for programming networks of devices forming an open dynamic topology with gossip-based communication. In the \ac{iiot} settings, this covers the devices at the {\em Far Edge} layer including in particular mobile devices, such as wearables carried by workers, or sensors attached to machines and objects that move, such as forklifts and pallets.

The main contributions of this paper are as follows:
\begin{enumerate}
\item
  we extend the FCPP library with the $\spawnK$ construct enabling {\em aggregate processes};
\item
  we port the extended FCPP library to a resource constrained, IoT physical DecaWave board, equipped with the Contiki NG operating system;
\item
  we identify several \ac{iiot} use cases that can be addressed by applying the \ac{ap} paradigm to program a network of devices equipped with FCPP;
\item
  we provide simulations of such \ac{iiot} use cases; and
\item
  we provide preliminary experiments with physical DecaWave boards running FCPP.
\end{enumerate}

The paper is structured as follows. Section \ref{sec-relwork} 
reviews related work. Section \ref{sec-ap} 
briefly recalls \ac{ap} and its FCCP implementation \cite{a:fcpp}, which is a library providing \ac{fc} as a C++ internal \ac{dsl}. Section \ref{sec-embedded} describes the fundamental step of porting the FCPP library on a DecaWave board, which provides wireless communication on a resource contrained System on Chip (SOC). Section \ref{sec-ap-iiot} links the powerful constructs of \ac{fc}  to some types of services that can be implemented in an \ac{iiot} setting.  Section \ref{sec-exp} describes an \ac{iiot} scenario of a \emph{Warehouse App}, an AP service run on hardware with the physical characteristics of the DecaWave board; then simulates its operation through the FCPP 3D simulator, and presents preliminary experiments with physical DecaWave boards running a port of the FCPP library. Finally, section \ref{sec-end} concludes the paper discussing further research directions.


\section{Related work}
\label{sec-relwork}

\subsection{Edge Computing in the \ac{iiot}}
Distributed edge architectures have been explored in the context of Data Management for the \ac{iiot}, which involves generating, aggregating, storing, analysing and requesting data. These are the main kinds of tasks we shall discuss in the present paper for \ac{ap} 
powered \ac{iiot} edge devices.

In \cite{urrehman:FGCS19}, the authors review the role of \ac{bda} in the \ac{iiot} setting. They recall that BDA involves several sub-tasks, such as data engineering, preparation, and analytics; and, in additiom, the management and automation of the data pipeline. They propose an architectural model called {\em Concentric Computing}, whereby the elements of a BDA/IIoT system are placed in concentric circles from the external far edge devices, to the outer and inner edge servers, to the cloud services.

According to Concentric Computing, computational and storage resources must be offered by different devices and systems across the complete set of cirles. The objectives are expressed in terms of storage, in-network data movement, energy consumption, privacy, security and real-time knowledge availability. Therefore, in many cases priority should be given to devices and systems near data sources to ensure real-time or near real-time intelligence near end points, IoT devices and other data sources in IIoT systems

In \cite{raptis:WIMOB17} the authors propose a \ac{dml} for the \ac{iiot} that is separated from the Network Routing. They focus on the lower ({\em far-edge}) layer of the \ac{iiot} architecture, and consider a typical scenario where a set of source nodes produce data, a set of destination nodes require those data, and a maximum latency $L_{max}$ is tolerated by destinations. They borrow the idea of {\em proxy} from other data distribution contexts (e.g., Content Distribution Networks \cite{chen2002scan} and Multimedia Streaming \cite{Wu:2004}), and dynamically compute a caching scheme on proxy nodes that aims at improving latency while keeping the number of proxies (and, thus, the data redundancy costs) as low as possible.

In \cite{rao:COMSNETS17}, the authors explicitly consider user \ac{ue} that features 5G connectivity as a means to fill the gap between the intelligence embedded in the tools, shop-floors, and conveyor belts and the cloud services able to process such information. The main problem becomes that of associating each IoT device to a \ac{ue} in a (approximately) optimal way. The authors find that the greedy association scheme, whereby the IoT devices associate with the \ac{ue}s with probabilities that are proportional to the number of devices that the \ac{ue}s can support, performs best. An interesting aspect of this study is the variation of several parameters during (simulated) tests, including the number of IoT devices, the number of \ac{ue}ss, the data arrival probability at the IoT devices, and the evolution of uplink data processes of the \ac{ue}s.

\subsection{Programming ensembles of devices spread over space at the far edge}

Different development approaches for systems involving a potentially vast number of heterogeneous devices that need to coordinate to perform collective tasks
by relying on
 proximity-based interactions (as in wireless sensor networks) have been proposed in literature. In the following we classify them into
 five categories, identified by  a survey~\cite{SpatialIGI2013}.

%
%
%
\begin{itemize}

\item \emph{Foundational approaches} propose compact formalizations aimed at modelling the interaction of groups in
 complex environments. Most of them extend  $\pi$-calculus~\cite{Milner:1992a}.
They include, for instance:  models of environment structure (from "ambients" to 3D abstractions) \cite{DBLP:conf/cie/CardelliG10,ambients,Milner200660};
shared-space abstractions allowing multiple processes to interact in a decoupled way \cite{klaim,VCMZ-TAAS2011}; and attribute-based models featuring declarative
specification of the target of communication for dynamically creating ensembles \cite{SCEL}.

\item \emph{Device abstraction languages} are aimed at allowing programmers to focus on cooperation and adaptation, by making the details of device interactions
 implicit. They include, for instance: TOTA~\cite{tota}, which supports programming tuples with reaction and diffusion rules;  the SAPERE approach~\cite{VPMSZ-SCP2015}, which supports similar rules
embedded in space and apply semantically; the $\sigma\tau$-Linda model~\cite{spatialcoord-coord2012}, which supports manipulation of tuples over space and time;
 MPI~\cite{MPI2}, which  allows to declaratively expresses topologies of processes in supercomputing applications; NetLogo~\cite{sklar2007netlogo}, which
 provides abstract means to interact with neighbours according to the cellular automata style; and Hood~\cite{hood}, which features  implicit sharing of values with neighbours.

\item \emph{Pattern languages} provide adaptive means for composing geometric and/or topological constructions, with little focus on
 computational capability.  They include, for instance: the Origami Shape Language~\cite{nagpalphd}, which allows to imperatively specify geometric folds that are compiled
  into processes identifying regions of space; the Growing Point Language~\cite{coorephd}, which  allows to describe topologies in terms of a ``botanical'' metaphor with
   growing points and tropisms; ASCAPE~\cite{inchiosa2002overcoming}, which supports agent communication by means of topological abstractions and a rule language; and a
    catalogue of self-organisation patterns \cite{FDMVA-NACO2013}, which organises a variety of mechanisms from low-level primitives to complex self-organization patterns.

\item \emph{Information movement languages} are the complement of pattern languages. They provide support summarising information obtained from across
 space-time regions of the environment and streaming these summaries to other regions, however, they provide  little control over the patterning of that computation.
They include, for instance: TinyDB~\cite{tinydb}, which views a wireless sensor network as a database; Regiment~\cite{regiment}, which uses a functional language to be compiled into
 protocols of device-to-device interaction; and KQML~\cite{Finin94kqml}, an  agent communication language.

\item \emph{Spatial computing languages}  provide flexible mechanisms and abstractions for spatial aspects of computation. They avoid  the limiting
 constraints of the other categories. They include, for instance: the Lisp-like functional language and simulator Proto~\cite{proto06a},
  for programming wireless sensor networks with the
  notion of computational fields; and the rule-based language MGS~\cite{GiavittoMGS02},  for computation of and on top of topological complexes.
\end{itemize}

As pointed out in~\cite{bpv:aggregate:programming}, the successes and failures
of the above languages, suggest that arraging
 adaptive mechanisms to be implicit helps to ensure simple and transarent  composition of aggregate-level modules and subsystems.
 This observation is further pursued by a recent survey~\cite{vbdacp:survey},  which overviews \ac{ap}
  and its foundation provided by the \ac{fc}.

\section{Aggregate Programming in FCPP}
\label{sec-ap}

\ac{ap}~\cite{bpv:aggregate:programming,vbdacp:survey} is an approach for programming networks of devices by abstracting away from individual devices behaviour and focusing on the global, aggregate behaviour of the collection of all devices.
It assumes only local communication between neighbour devices, and it is robust with respect to devices joining/leaving the network, or failing (open dynamic topology).
Beside communicating with neighbours, the devices are capable to perform asynchronous computations. In particular, every device performs periodically the same sequence of operations,
with an usually steady rate:
\begin{enumerate}
\item collection of received messages,
\item computation of a program that is the same for all the devices, and
\item transmission of messages
\end{enumerate}

\ac{ap} is formally backed by \ac{fc}~\cite{avdpb:TOCLhfc}, a small functional language for expressing aggregate programs,
which currently has three incarnations as a full-fledged \ac{dsl}: the Scala internal \ac{dsl}/library \emph{ScaFi (Scala Fields)}~\cite{CasadeiVAD20}, the Java external \ac{dsl} \emph{Protelis}~\cite{pvb:protelis},
and the C++ internal \ac{dsl}/library  \emph{FCPP}~\cite{a:fcpp}.
In this paper we focus on the FCPP incarnation, because it is the only one that lends itself to be ported to devices with constrained resources, such as the ones we consider for \ac{iiot} (see Section \ref{sec-embedded}).

FCPP  is based on an extensible software architecture, at the core of which are {\em components}, that define abstractions for single devices ({\em node}) and overall network  orchestration ({\em net}), the latter one being crucial in simulations and cloud-oriented applications. In an FCPP application, the two types {\em node} and {\em net} are obtained by combining a chosen sequence of components, providing the needed functionalities in a mixin-like fashion.

\begin{figure}[tbp]
\centerline{\small\framebox[\linewidth]{$
	\begin{array}{@{\hspace{0pt}}l@{\hspace{0pt}}c@{\hspace{1pt}}l @{\hspace{40pt}} l@{\hspace{2pt}}c@{\hspace{2pt}}l@{\hspace{0pt}}}
		\multicolumn{6}{@{\hspace{0pt}}l@{\hspace{0pt}}}{\mbox{aggregate function declaration}} \\[3pt]
		\FUNCTION & \BNFcce & \multicolumn{4}{@{\hspace{2pt}}l@{\hspace{0pt}}}{\defK ~ \type ~ \dname(\argK, \type ~ \xname*) ~ \{ \codeK ~ \retK ~ \e; \}} \\[3pt]
		\hline \\[-5pt]
		\multicolumn{6}{@{\hspace{0pt}}l@{\hspace{0pt}}}{\mbox{aggregate expression}} \\[3pt]
		\e & \BNFcce & \multicolumn{4}{@{\hspace{0pt}}l@{\hspace{0pt}}}{\xname \BNFmid \lvalue \BNFmid \type\{\e*\} \BNFmid \uname \e \BNFmid \e ~\oname~ \e \BNFmid \pname(\e*) \BNFmid \nodeK.\cname(\e*) \BNFmid \fname(\callK, \e*)} \\[3pt]
		& & \multicolumn{4}{l}{\BNFmid \type ~ \xname = \e; ~ \e \BNFmid [\&](\type ~ \xname*) \toK \type \, \{ \retK ~ \e; \} \BNFmid \e ~?~ \e : \e} \\[3pt]
		\hline \\[-5pt]
		\multicolumn{3}{@{\hspace{0pt}}l@{\hspace{0pt}}}{\mbox{type}} &
		\multicolumn{3}{@{\hspace{0pt}}l@{\hspace{0pt}}}{\mbox{aggregate function}} \\[3pt]
		\type & \BNFcce & \tvar \BNFmid \btype \BNFmid \ttype\ap{\type*, \lvalue*} &
		\fname & \BNFcce & \bname \BNFmid \dname \\[3pt]
		\hline \\[-5pt]
		\multicolumn{6}{@{\hspace{0pt}}l@{\hspace{0pt}}}{\mbox{built-in aggregate functions}} \\[3pt]
		\bname & \BNFcce & \multicolumn{4}{@{\hspace{2pt}}l@{\hspace{0pt}}}{\mathtt{old} \BNFmid \mathtt{nbr} \BNFmid \mathtt{spawn} \BNFmid \mathtt{self} \BNFmid \mathtt{mod\_self} \BNFmid \mathtt{map\_hood} \BNFmid \mathtt{fold\_hood} \BNFmid \mathtt{mux}
                }
	\end{array}
$}}
\caption{Syntax of FCPP aggregate functions.} \label{fig:syntax}
\end{figure}

Compared to the original presentation of FCPP~\cite{a:fcpp}, we have added a fundamental construct for supporting \emph{aggregate processes} \cite{casadei2019aggregate}, namely the built-in {\em spawn} function (see below).
Aggregate processes can be figured as {\em computational bubbles} that involve a subset of the devices running a given FCPP program; such bubbles can spring out, expand, perform some work, stretch and vanish.
Given a computational bubble, a device can be either within the bubble (i.e., participating in the computation and bubble spreading), external to the bubble (i.e., not participating in the computation), or at the border of the bubble (i.e., participating in the computation but not in bubble spreading).

A fundamental aspect is that every process instance is automatically propagated by all the participating (internal) devices to their neighbours. Therefore, when a device generates a process, it just needs not to leave it immediately in order to propagate it to its neighbours; and so on. Unless nodes explicitly indicate that they are willing to leave the process, the process will tend to expand to every reachable node. On the other hand, when a device leaves a process, even if it happens to be the process creator, it is up to the other nodes still in the process to decide whether they also want to leave (eventually leading up to the termination of the whole process) or not. It is also possible, however, to explicitly initiate a propagating shutdown of the process (through a special status, see below).

The syntax of aggregate functions in FCPP is given in Fig.~\ref{fig:syntax}. It should be noted that, since FCPP is a C++ library providing an internal \ac{dsl}, \emph{an FCCP program is a C++ program}
 (so all the features of C++ are available). We use $*$ to indicate an element that may be repeated multiple times (possibly zero).

An \emph{aggregate function declaration} consists of keyword $\defK$, followed by the return type $\type$ and the function name $\dname$, followed by a parenthesized sequence of comma-separated arguments $\type ~ \xname$ (prepended by the keyword $\argK$), followed by an \emph{aggregate expression} $\e$ (within brackets and keywords $\codeK ~ \retK$).
\emph{Aggregate expressions} can be either:
\begin{itemize}
	\item a \emph{variable} identifier $\xname$, or a C++ \emph{literal value} $\ell$ (e.g.~an integer or floating-point number);
	\item an \emph{object} of type $\type$ built through a class constructor call $\type\{\e*\}$ with arguments $\e$;
	\item an \emph{unary operator} $\uname$ (e.g.~$-$, $\sim$, $!$, etc.) applied to $\e$, or a \emph{binary operator} $\e ~\oname ~ \e$ (e.g.~$+$, $*$, etc.);
	\item a \emph{pure function call} $\pname(\e\ast)$, where $\pname$ is a basic C++ function which does not depend on node information nor message exchanges
	\item a \emph{component function call} $\nodeK.\cname(\e\ast)$, where $\cname$ is a function provided by some component, depending on node information but not on messages;
	\item an \emph{aggregate function call} $\fname(\callK, \e*)$, where $\fname$ can be either a defined aggregate function name $\dname$ or an aggregate built-in function $\bname$ (see below);
	\item a \emph{let-style statement} $\type ~ \xname = \e_1; ~ \e_2$, declaring a variable $\xname$ of type $\type$ with value $\e_1$ referable in $\e_2$;
	\item a \emph{conditional branching} expression $\e_\text{guard} ~?~ \e_\top : \e_\bot$, such that $\e_\top$ is evaluated and returned if $\e_\text{guard}$ evaluates to \lstinline|true|, while $\e_\bot$ is evaluated and returned if $\e_\text{guard}$ evaluates to \lstinline|false|.
\end{itemize}
The \lstinline|old| and \lstinline|nbr| built-in functions, constitute the fundamental constructs of Field Calculus; in FCPP, they are overloaded to several different signatures:
\begin{itemize}
	\item
	  $\oldK(\callK, v_0, v)$ with $v_0, v$ of type $\type$ returns the value fed as second argument $v$ in the \emph{previous round} of computation (thus introducing one round of delay), defaulting to $v_0$ if no such value is available;

    \item
      $\oldK(\callK, v)$ is a shorthand for $\oldK(\callK, v, v)$;

	\item
      $\oldK(\callK, v_0, f)$ computes the result of applying $f$ to the value of the whole $\oldK$ function at the previous computation cycle (using $v_0$ if no such value is available);

    \item
	  $\nbrK(\callK, v_0, v)$ with $v_0, v$ of type $\type$ returns the \emph{neighbouring field} of values fed as second argument $v$ in the previous round of computation of \emph{neighbour nodes}, defaulting to $v_0$ for the current node if no such value is available for it;

    \item
	$\nbrK(\callK, v)$ is a shorthand for $\nbrK(\callK, v, v)$;

    \item
	  $\nbrK(\callK, v_0, f)$ (whose logic is described in \cite{abdpv:share} as the {\em share} operator), computes the result of applying $f$ to the neighbouring field of values of the whole $\nbrK$ function at the previous computation cycle of neighbour nodes (using $v_0$ for the current node if no such value is available).
\end{itemize}

The newly implemented \lstinline|spawn| built-in function has the following signature:
\[\spawnK(\callK, p, ks, v_0, ...)\]
and spawns an \emph{aggregate process} corresponding to function $p$ for every \emph{key} in the container $ks$, passing the \emph{values} of the (possibly empty) sequence $v_0, \ldots$ as further input to each of them.
The aggregate process function $p$ takes as arguments a key and a sequence of values, and returns a pair consisting of a result and a process status. Currently, FCPP supports overloads of $\spawnK$ for two different types of process status:
\begin{enumerate}
\item $\mathtt{status}$, that is one of the following constants:
  \begin{enumerate}
  \item $\mathtt{terminated}$: the node wants to shutdown the computation (propagating to its neighbours)
  \item $\mathtt{external}$: the node is not part of the computation
  \item $\mathtt{border}$: the node is at the border of the computation (see above)
  \item $\mathtt{internal}$: the node is within the computational bubble
  \item $\mathtt{*\_output}$ (where $\mathtt{*}$ is one of $\mathtt{terminated}$, $\mathtt{external}$, $\mathtt{border}$, or $\mathtt{internal}$): the node is in status $\mathtt{*}$ and the output of function $p$ should be returned by $\spawnK$
  \end{enumerate}
\item $\mathtt{bool}$, with $\truevalue$ and $\falsevalue$ corresponding to the $\mathtt{internal\_output}$ and $\mathtt{border\_output}$ values of type $\mathtt{status}$
\end{enumerate}
The $\spawnK$ function itself returns an unordered map from the keys of the processes with an $\mathtt{*\_output}$ state to their output values, so that such output values can be used for further computations in the current round.

The other built-in aggregate functions currently available are:
\begin{itemize}
	\item $\mathtt{self}(\callK, \phi)$, which given a value $\phi$ of $\ftype{\type}$ type returns the value $\phi(i)$ taken by the neighbouring field $\phi$ for the current node (of identifier $i =$ \lstinline|node.uid|);
	\item $\mathtt{mod\_self}(\callK, \phi, v)$, which given a value $\phi$ of $\ftype{\type}$ type, returns the same value with $\phi(i)$ changed to $v$, where $i =$ \lstinline|node.uid|;
	\item $\mathtt{map\_hood}(\callK, f, v*)$ which applies $f$ point-wise to a sequence of local or field values $v*$;
	\item $\mathtt{fold\_hood}(\callK, f, \phi)$ which \emph{folds} the values in the range of $\phi$ of $\ftype{\type}$ type through the commutative and associative binary operator $f$ of type $(\type,\type) \toK \type$, reducing them to a single value of type $\type$;\footnote{In Field Calculus, a neighbouring field always has at least a value for the current node; thus, folding is well-defined.}
	\item $\mathtt{fold\_hood}(\callK, f, \phi, v)$ which folds $\phi$ as above, using $v$ instead of the value of $\phi$ for the current device: in other words, it is equivalent to $\mathtt{fold\_hood}(\callK, f, \mathtt{mod\_self}(\callK, \phi, v)$;
        \item $\mathtt{mux}(\callK, e_c, e_t, e_f)$, which evaluates all the expressions $e_c, e_t, e_f$ and returns the value of either $e_t$ or $e_f$ based on the Boolean value of $e_c$; note how $\mathtt{mux}$ differs from the conditional branching expression described above which evaluates only the branch selected by the condition.
\end{itemize}

\section{Port of FCPP on a DecaWave Board}
\label{sec-embedded}
A fundamental step for using FCPP in real-world \ac{iiot} scenarios is  porting it to a suitable platform, including hardware and operating system. We have chosen the DWM1001C module produced by Decawave, which is currently used by Reply\footnote{The company which co-operated to the present study: \url{https://www.reply.com}.} to offer solutions to some of its industrial customers.

The DWM1001C module integrates the Nordic Semiconductor nRF52832 general-purpose system on a chip (SoC), the Decawave DW1000 Ultra Wideband (UWB) transceiver and the STM  LIS2DH12TR 3-axis accelerometer.
The nRF52832 SoC offers a 64MHz ARM Cortex-M4 CPU with floating-point unit, a Bluetooth Low Energy (BLE) transceiver with Bluetooth 5 support, a 512 KB flash memory and a 64 KB RAM. Decawave also offers a developer board (DWM1001-DEV) with a battery connector, a charging circuit, eight LEDs, two buttons, a USB connector and a J-Link on board debug probe for debugging and logging.

The UWB transceiver allows the module to perform communication over an higher range than the BLE transceiver and to compute the distance between two modules ({\em ranging}) with a precision of up to 10 centimeters.
In particular, in our experiments we measured a range of communication in open air of around 70 meters with UWB and of around 25 meters with BLE, with a ranging error of under 30 centimeters with UWB.
We also measured the energy consumption by the UWB transceiver to be around three times higher than the BLE transceiver.

We based our FCPP porting on the existing work of the D3S Research Group of the University of Trento\footnote{See  \url{https://github.com/d3s-trento/contiki-uwb}.}, which includes a port of the Contiki operating system\footnote{See  \url{https://github.com/contiki-os/contiki}.} on the DWM1001C and a UWB driver with ranging API.
Contiki is an open source operating system for microcontrollers in the IoT, which provides high level APIs for cooperatives threads (proto-threads, see \cite{Dunkels:2006}), timers and networking (with protocols for each network stack layer).
We upgraded the porting to the newer Contiki NG,\footnote{See \url{https://github.com/contiki-ng/contiki-ng}.} which provides a partial support for the C++ programming language required by FCPP, and
we improved the C++ support by integrating the C++ clock API and standard output with the Decawave hardware.
Contiki NG offers a low code footprint of just about 100kB and the possibility to configure memory usage to be as low as 10kB. The access to the UWB and BLE features of DWM1001C is granted, respectively, by a port of the UWB driver for Contiki developed by the D3S Research Group (see above), and by the Soft Device for BLE offered by the Nordic SDK for the nRF52832 SoC.

In order to connect FCPP with the UWB and BLE drivers, we had to extend it. In particular, the $\mathtt{hardware\_connector}$ component included in the library handles the communications between physical (hardware) nodes. The constructor of its $\mathtt{node}$ class receives an object whose type is the class implementing the communication functions on a specific hardware. For our present purposes, we have created two classes (FCPP drivers), one for UWB and one for BLE. In order to work with the $\mathtt{hardware\_connector}$ component, they just need to expose:
\begin{itemize}
\item a constructor taking a $\mathtt{node}$ which represents the current device;
\item a constructor taking a $\mathtt{node}$ and a data structure for configuring the driver;
\item a $\mathtt{send}$ method taking a vector of characters to send; and
\item a $\mathtt{receive}$ method that returns a vector of messages from neighbours.
\end{itemize}

The UWB FCPP driver exploits the native UWB driver to broadcast all communications between devices over the CSMA protocol with the UWB transceiver, reaching a greater distance of communication compared to BLE at the expense of an higher energy consumption. The UWB configuration allows to send messages of size up to 116 bytes.

The BLE FCPP driver exploits the native BLE driver to broadcast the messages using the Bluetooth 5 extended advertisement. Moreover, it uses the UWB transceiver to perform the ranging between devices for computing their distances. The main goal of the protocol for intermixing BLE communication with ranging is that of keeping the UWB transceiver in sleep mode for as long as possible to reduce energy consumption. More precisely, the main steps are as follows:
\begin{itemize}
\item when a node sends a message through the BLE native driver, it adds a prefix with a list of neighbour nodes it wants to invite to do a ranging session at the beginning of the next round;
\item the node then prepares to do ranging with the neighbours in the list at the beginning of its next round; and
\item when a node receives a message through the BLE native driver, and it appears in the prefix list, it prepares to do ranging with the sender at the beginning of the next round of the sender.
\end{itemize}
Thanks to the synchronization information exchanged by piggybacking the BLE messages, the nodes can turn on their UWB transceivers just for the time needed for performing the ranging operations.
The BLE configuration allows to send messages of over 200 bytes, with the actual size depending on the ranging configuration, i.e., on the maximum size of the prefix devoted to synchronize ranging, which is not available for the regular payload.

As mentioned above, a serious constraint of the DWM1001C module is the quantity of RAM, limited to 64kB. Thanks to the small footprint of Contiki NG and of FCPP, it has been possible to leave approximately 16kB of stack space and 16kB of heap space to be used by applications built on FCPP.

\section{\ac{ap} Services in an Industrial IoT}
\label{sec-ap-iiot}

\subsection{Overall Hardware Architecture}

\begin{figure}[t]
	\centering
	\includegraphics[width=0.75\textwidth]{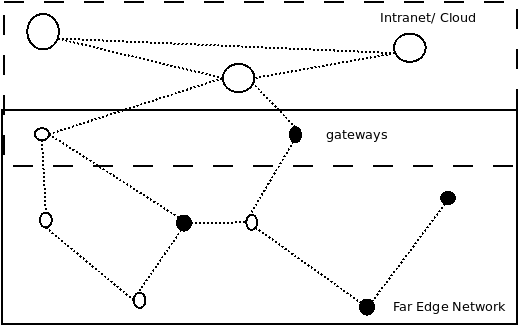}
	\vspace{-10pt}
	\caption{Schema of the Architecture for the \ac{iiot} Service Scenarios.} \label{fig:arch}
\end{figure}

Figure~\ref{fig:arch} shows the schematic architecture that we assume for the \ac{iiot} scenarios we address. It is derived from the one in Figure \ref{fig:iiot-arch}, but is specifically tailored to the assumptions and focus of this paper.

We assume two networks, based on different technologies and covering different layers:
\begin{enumerate}
\item a {\em Far Edge net} $\edgenet$ composed of low-end computational nodes that communicate point-2-point based on proximity. Corresponds to the {\em Edge} layer in Figure \ref{fig:iiot-arch} but assumes specifically point-2-point wireless communication; and
\item an {\em Intranet and Cloud net} $\intnet$ that connects more powerful systems within the factory among them and, possibly, with external cloud systems, using standard internet technologies and protocols. Corresponds to the {\em Network} and {\em Cloud} layers in Figure \ref{fig:iiot-arch}, that we do not need to distinguish for our purposes.
\end{enumerate}

We envision the following types of computational nodes:
\begin{enumerate}
\item {\em fixed} nodes $\fixn$ of the $\edgenet$ net, associated with fixed equipment such as machines, controllers, or fixed sensors (depicted as empty circles in Figure \ref{fig:arch});
\item {\em mobile} nodes $\mobn$ of the $\edgenet$ net, associated with mobile equipment (e.g., forklifts) and users (depicted as full circles in Figure \ref{fig:arch});
\item {\em gateway} nodes $\gwyn$ that are nodes that can communicate with both the $\edgenet$ nodes and the $\intnet$ nodes, thus making it possible to route data between the two nets; and
\item {\em cloud} central system nodes $\clsys$ that can provide: long term storage; expensive data analysis and ML; connection to a Digital Twin; etc.
\end{enumerate}
It should be noted that the set of nodes $\mobn$ can change dynamically quite often, since, e.g., users enter/leave the factory or moving machines are switched on/off. Moreover, gateway nodes $g \in \gwyn$ may be either fixed or mobile.

Sensors and actuators can be associated with both mobile and fixed nodes. For convenience, we spell two important categories:
\begin{enumerate}
\item {\em distance} sensors $\dists$, that can detect the distance between nodes of the $\edgenet$; and
\item {\em parameter} sensors $\params$, that measure the values of relevant parameters at their location.
\end{enumerate}
We do not detail how the nodes of $\edgenet$ communicate with their associated sensors and actuators: they may, e.g., be directly connected through a common physical board, or use a short-range wireless technology such as BLE. We just assume that $\edgenet$ can retrieve data from associated sensors and issue commands to associated actuators.

\subsection{Services Software Architecture}
\label{ssec-ap-iiot-swarch}

\ac{ap}-based services consist of the execution of FCPP programs by the nodes of the $\edgenet$.
An important role, however, is played by Aggregate Processes and how they are generated, managed, and terminated.

Consider a node $n_c$ ({\em client}) that must reach another node $n_s$ ({\em server}) for receiving a service or an information. A simple schema would be to have $n_c$ broadcast its request into the $\edgenet$, then collect the replies from the available service providers $n'_s, n''_s, \ldots$, and finally choosing to adopt, e.g., the reply of the closest one, and discard the others.
The problem with this schema, is that, in order for it to work, every node in the $\edgenet$ should be executing an FCPP program that includes the logic for the broadcast from $n_c$ and the generation and collection of replies from service provider nodes. This is obviously not practical, since the client can be any node in $\edgenet$ and there may be many possible types of requests that must be handled and answered according to different logics.

\begin{figure}[t]
	\centering
	\includegraphics[width=0.8\linewidth]{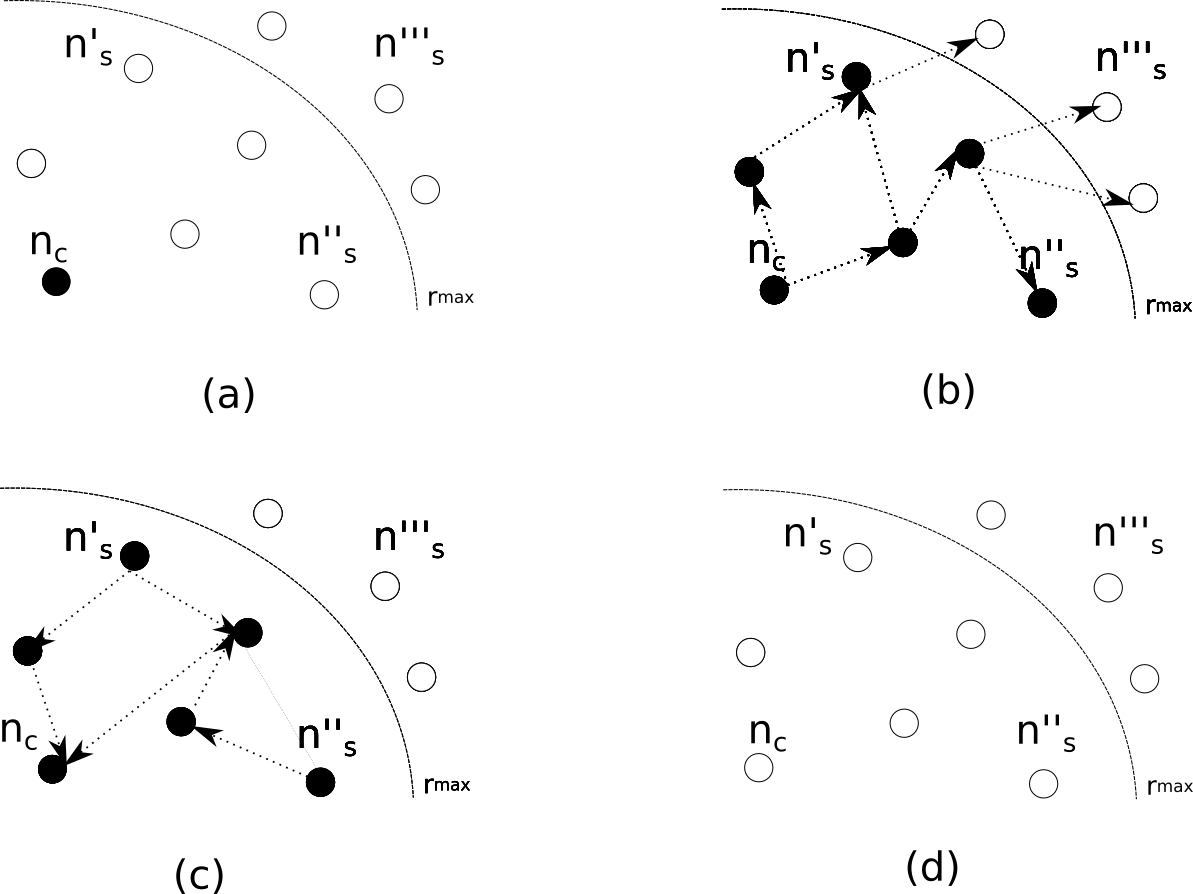}
	\caption{Life of an aggregate process spawned by a  client node (a-d). See explanation in the text.} \label{fig:aggproc}
\end{figure}

This is where aggregate processes come into play as a fundamental building block of  \ac{ap} systems. In a situation like the one described above:
\begin{itemize}
\item The {\em client} node $n_c$ creates a process, by passing a suitable key (e.g., its node id $id_c$) to a $\spawnK$ that appears in the FCPP program being executed:
  \[\spawnK(\callK, p, \{id_c\}, v_0, ...)\]
  Note that, up to the current round the $\spawnK$ has always been executed by node $n_c$ as part of the program, but since it was given an empty set of keys, it immediately returned as a no-operation (if it didn't execute processes generated by other clients).
  Figure \ref{fig:aggproc} (a) depicts $n_c$ as a full circle to indicate that it participates to the new process.
\item The function executed by the new process is the one specified as the $p$ argument to $\spawnK$, and of course depends on how exactly the interaction between $n_c$ and the service provider(s) should happen; it receives the key of the process $id_c$ and the additional parameters $v_0, ...$.
\item Typically, $n_c$ wants to place a limit $r_{max}$ to the maximum radius of the process expansion, and this can be achieved by passing $r_{max}$ (shown as a dotted arc in Figure \ref{fig:aggproc}) as one of the additional parameters $v_0, ...$ of $\spawnK$ that are forwarded to $p$.
\item The process created by $n_c$ starts to spread automatically to the neighbours of $n_c$, and on to more and more nodes $n_p \in \edgenet$.
  Figure \ref{fig:aggproc} (b) shows as full circles the nodes within $r_{max}$ to which the process propagates.
\item Consider a node $n_p$ to which the process has just been propagated:
  \begin{itemize}
  \item $n_p$ executes the $p$ function, which, exactly as in $n_c$, receives the key of the process $id_c$ and the other parameters including $r_{max}$.
  \item The body of function $p$ should be such that node $n_p$ gets to know whether it is beyond the maximum radius $r_{max}$ from node $n_c$ (e.g., by participating to a {\em gradient} aggregate computation, see section \ref{ssec-share}).
  \item If so, $n_p$ should immediately leave the process (i.e., return an $\mathtt{external}$ or $\mathtt{false}$ status).
    In Figure \ref{fig:aggproc} (b), the nodes beyond $r_{max}$ remain empty circles although the nodes within the bubble try to propagate the process to them.
  \item Otherwise, it should determine whether it can (try to) provide the service requested by $n_c$; if yes, it should send its reply to $n_p$ (e.g., by participating as a producer to a {\em collection} aggregate computation, see section \ref{ssec-share}).
    Figure \ref{fig:aggproc} (c) shows the replies originating from the server nodes $n'_s, n''_s$ flowing towards $n_c$ through the $\edgenet$.
  \item When $n_c$ decides to terminate the process, it initiates the shutdown by returning a $\mathtt{terminated}$ status, which propagates to the other nodes within the bubble.
    Figure \ref{fig:aggproc} (d) depicts all the nodes, including $n_c$, as empty circles to indicate that they have all left the process.
  \end{itemize}
\end{itemize}

\subsection{Safety Services}

An important set of \ac{ap}-based services that can be implemented in the architecture of Figure~\ref{fig:arch} are those that improve the safety of the industrial environment and, in particular, of the people that populate it at any given time.

The typical schema for the \ac{ap} implementation of a safety service is the following:
\begin{enumerate}
\item let $\nssafe$ be the subset of nodes of the $\edgenet$ net involved in the safety service;
\item nodes $\nssafe$ continuously exchange relevant information (i.e., their velocity) with neighbours, and receive data from their sensors;
\item in particular, if the node is equipped with a distance sensor $s \in \dists$, the built-in funtion $\mathtt{nbr\_dist}$ returns a field of type \lstinline|field<real_t>| associating each neighbour UID (of type \lstinline|device_t|) with its distance from the current node (of type \lstinline|real_t|);
\item the sensor data is evaluated w.r.t. the internal status and with the safety-related data collected from the neighbours;
\item if a danger is detected, the node can react immediately (e.g., stop moving) and initiate an information sharing process to share the detected danger with other
$\edgenet$ and/or $\intnet$ nodes (see sections \ref{ssec-share}, \ref{ssec-cloud}).
\end{enumerate}

It should be noted that even a simple safety process as the one just described can take advantage of the intelligence provided by \ac{ap}. For example, the safety threshold of the distance between two nodes can depend on whether the two nodes are not moving, moving towards each other, or moving away from each other. This can be easily achieved with \ac{ap}.

\subsection{Information Sharing for Intelligent Decisions}
\label{ssec-share}

The sharing of information among $\edgenet$ nodes (without involving the $\intnet$ nodes) can provide many useful, robust and low-latency services.
In particular, when executed as an Aggregate Process, it can support the request of a service by a node and the collection of the replies from potential servers, as explained in section \ref{ssec-ap-iiot-swarch}.

Information sharing can typically follow two alternative ways: the first one assumes that the nodes that directly sense (or, more generally, own) relevant information {\em spread} it to the rest of the net; the second one assumes that nodes that need information, {\em collect} it from the other nodes possibly {\em accumulating} it into a suitable data structure. The spreading and collection of information are fundamental blocks of the \ac{ap} paradigm, and are sometimes named, respectively, $G$-block and $C$-block in the literature \cite{viroli:selfstabilisation}.
Referring again to the service request/reply described in section \ref{ssec-ap-iiot-swarch}, the client node can spread its request with a $G$-block, and collect the replies with a $C$-block.

The typical schema for the \ac{ap} implementation of information spreading in our reference setting is the following:
\begin{enumerate}
\item let $\nscons$ be the subset of nodes of the $\edgenet$ net interested in the information produced by a node $\prodnode$
\item node $\prodnode$ produces the information and broadcasts it (see below) to all the $\edgenet$ nodes
\item the nodes outside $\nscons$ are used just to propagate the information, while the nodes in $\nscons$ also pick-it up and use it for their computations, possibly including making decisions, and posting a reply
\end{enumerate}

\begin{figure}[t]
\centering
{\small
\begin{lstlisting}[]
GEN(P,T) T broadcast(ARGS, P const& distance, T const& value) { CODE
  return nbr(CALL, value, [&] (field<T> x) {
    return get<1>(min_hood(CALL,
                           make_tuple(nbr(CALL, distance), x),
                           make_tuple(distance, value)));
  });
}
\end{lstlisting}
}
\caption{The broadcast function implementation in FCPP.} \label{fig:broadcast}
\end{figure}

Figure \ref{fig:broadcast} shows the implementation of the \lstinline|broadcast| function in FCPP.
First of all, we note that it is a templated function, which parametrizes the types \lstinline|P|, \lstinline|T| of the \lstinline|distance| and \lstinline|value| arguments: in this case, the $\defK$ keyword is substituted with the alternative $\mathtt{GEN}(T^\ast)$ listing the parameter types. The \lstinline|distance| parameter represents the (estimated) distance of the node from the source of the information, i.e. it is $0$ for the node that generates the information, and can be estimated with a suitable aggregate function for the other nodes in $\edgenet$. The FCPP library offers several functions for the distance estimation, ranging from a basic function \lstinline|abf_hops| counting the number of hops from the source with the Bellman-Ford algorithm, to sophisticated functions such as \lstinline|bis_distance| \cite{audrito:bisgradient}, or \lstinline|flex_distance| \cite{Beal:FLEX}. Clearly, a more accurate estimate can be achieved if the underlying system provides $\dists$ sensors for measuring the distance between each node and its neighbours.

The function body uses the form of \lstinline|nbr| that takes a lambda function that processes the field \lstinline|x| of most-recent values exchanged with neighbours to derive the new value for the current node. The lambda applies \lstinline|min_hood| to pairs $(d,v)$ for each neighbour $\delta$, where $d$ is the distance of $\delta$ from the source and $v$ is the value held by $\delta$; the result of \lstinline|min_hood| is thus a pair $(d',v')$ corresponding to the neighbour closest to the source. Finally, the use of \lstinline|get| extracts the value $v'$, and such a value is returned by \lstinline|broadcast| and will be associated with the current node in the field \lstinline|x| when \lstinline|broadcast| is computed again in the next round.

The \ac{ap} implementation of information collection and accumulation in our reference setting typically has the following schema:
\begin{enumerate}
\item let $\nsprod$ be the subset of nodes of the $\edgenet$ net involved in the production of information that must be collected by a node $\consnode$;
\item each node in $\nsprod$ produces a piece of information and aggregates it into the data that is collected towards $\consnode$ through the other $\edgenet$
 nodes (see below);
\item when the aggregated data reaches $\consnode$ it is picked-up and used.
\end{enumerate}
Note that, in the second step, it is possible to aggregate the information while it flows from the $\nsprod$ nodes to the $\consnode$. For instance, imagine that a client node $n_c$ has spawned a process to ask for a service, and it then collects the answers of servers $n'_s, n''_s, \ldots$ consisting of pairs \lstinline|pair<real,real>| where the second element is the value of the reply, and the first element is the degree of confidence that the server had in that reply. As the replies flow towards $n_c$, each node in $\edgenet$ will propagate only the pair with the highest first element. In this case, the aggregation function is thus the $max$ function, but any other aggregation function could be used in the process.

\begin{figure}[t]
\raggedleft
{\small
\begin{lstlisting}[]
GEN(P, T, U, G)
T sp_collection(ARGS, P const& distance, T const& value,
                        U const& null, G&& accumulate) { CODE
  return nbr(CALL, (T)null, [&](field<T> x){
    device_t parent = get<1>(min_hood(CALL,
                                      make_tuple(nbr(CALL, distance),
                                                 nbr_uid(CALL)) ));
    return fold_hood(CALL, accumulate,
                           mux(nbr(CALL, parent) == node.uid, x, (T)null),
                           value);
  });
}
\end{lstlisting}
}
\caption{The single-path collection function implementation in FCPP.} \label{fig:collect}
\end{figure}

Figure \ref{fig:collect} shows the implementation of the \lstinline|sp_collect| function in FCPP. The FCPP library also offers more sophisticated multi-path collection functions, namely \lstinline|mp_collect| and \lstinline|wmp_collect| \cite{a:aamas:weighted}, but the simpler \lstinline|sp_collect| serves well our current explanation purposes.
The \lstinline|sp_collect| templated function has the following type parameters: \lstinline|P|, of the \lstinline|distance| parameter, representing the (estimated) distance of the node from the consumer of the information; \lstinline|T|, of the \lstinline|value| parameter, representing the aggregate value awaited by the consumer; \lstinline|U|, of the \lstinline|null| parameter, the identity element of the accumuation function; and \lstinline|G|, of the \lstinline|accumulate| parameter, representing the accumulation function that should take two \lstinline|T| values and aggregate them into a a single output \lstinline|T| value.

Similarly to \lstinline|broadcast|, the function body uses the form of \lstinline|nbr| that takes a lambda function that processes the field \lstinline|x| of most-recent values exchanged with neighbours to derive the new value for the current node. First, the lambda applies \lstinline|get<1>| to the result of a \lstinline|min_hood| to get the node \lstinline|uid| of the neighbour closest to the consumer; we store the result in a \lstinline|parent| variable, to stress the fact that the flow of information follows a tree\footnote{It is easy to see that such a tree is a Single Source Shortest Path (SSSP) tree with the consumer as source, on the (unweighted) network graph.} from the furthest nodes to the consumer itself.

Then, the \lstinline|accumulate| function is exploited by \lstinline|fold_hood| to aggregate a field of \lstinline|T| values into a single result of type \lstinline|T|. Such a value is returned by \lstinline|sp_collection| and will be associated with the current node in the field \lstinline|x| when \lstinline|sp_collection| is computed again in the next round. Using the \lstinline|mux| function, the field aggregated by \lstinline|fold_hood| associates to each neighbour $\delta'$ of the current node $\delta$ a \lstinline|T| value as follows: if $\delta'$ has determined that $\delta$ is its \lstinline|parent|, the most recent value sent by $\delta'$ (contained in the \lstinline|x| field); otherwise the identity element \lstinline|null| of \lstinline|accumulate|. In other words, the current device $\delta$ aggregates all and only the values received from neighbours that chose it as their parent.

\subsection{Interaction with the Cloud}
\label{ssec-cloud}

The sharing of information between the $\edgenet$ nodes and the $\intnet$ nodes, and especially between the edge and the $\clsys$ nodes of the $\intnet$ that host cloud services, is different than the $\edgenet$-level sharing described in the previous section in the following main respects:
\begin{itemize}
\item all the communications must necessarily go through the gateway nodes $\gwyn$;
\item the amount of data collected by cloud nodes can be much higher, in general, than that required by services provided completely within the $\edgenet$;
\item typical benefits of the \ac{ap} paradigm such as low latency, robustness, and privacy, which apply to the $\edgenet$ services, may not apply to services that also require information to cross the $\intnet$.
\end{itemize}
Given these characteristics, the following additional mechanisms can be suitable:
\begin{itemize}
\item distribute the workload among $\gwyn$ gateway nodes as far as possible;
\item ensure some redundancy in the data transmitted to the $\clsys$ cloud nodes (for improving both latency and robustness).
\end{itemize}
In FCPP, the distribution of workload can be easily done by partitioning the nodes in $\edgenet$ in as many regions as there are gateway nodes in $\gwyn$. A design pattern specifically created to achive these goals in a fully distributed fashion through \ac{fc} itself is the SCR (Self-Organising Coordination Regions) pattern described in \cite{Pianini:FGCS21}.
The pattern supports distributed selection of the leaders, but in case all the $\gwyn$ nodes are used, the selection becomes trivial. Also the formation of the regions can be trivial in its simplest form, whereby each node decides to belong to the region associated with the the closest $\gwyn$ node.

As for ensuring redundancy, a natural approach is to extend the SCR pattern in such a way that regions overlap, i.e., each node belongs to two or more regions.
Again, the criteria adopted by a node to select the regions to join can have varying degrees of complexity depending on the context. If the $\gwyn$ nodes can be partitioned a-priori into two or more subgroups or {\em types}, a simple approach consists of each node joining one region per type.

\section{Experimental Validation}
\label{sec-exp}

\subsection{Case Study: Warehouse App}

To validate the proposed approach experimentally, we consider a scenario inspired by use-cases currently being investigated in Reply of smart warehouse management. We assume that warehouse workers move around a series of aisles with forklifts moving at a maximum speed of 10 km/h (a standard for forklifts). Pallets containing goods are arranged in a regular grid along aisles, while some empty pallets are available in a common loading zone, where every load and unload operation is performed. We assume that the warehouse is managed with a high turnover, so that goods are placed as close as possible to the relevant point of operation for them, without a fixed placement based on the good type. High turnover allows for greater efficiency in principle, but it also suffers from performance degradation as the warehouse starts to fill up: workers may need to perform long searches for a required good, or even to find an empty space for a new pallet. As a byproduct, a digital representation of the warehouse status (e.g., a digital twin) is usually inaccurate or impossible. Furthermore, workers may occasionally run into each other at aisle joints, damaging goods and slowing down the warehouse operations.

In order to overcome these issues, we propose a warehouse management app realising the following services:
\begin{enumerate}
	\item \emph{preventing accidental collisions}, by warning workers whenever another forklift is approaching with a speed greater than a threshold, within a given safety radius;
	\item \emph{providing route information} towards either empty spaces and goods matching a given query, presented to the interested warehouse worker by turning on led lights on neighbouring smart devices;
	\item \emph{collecting logs} of relevant events (loading/unloading of  goods and collision warnings) towards central points that are connected to the cloud.
\end{enumerate}
We assume that the app runs on a network of DWM1001C modules, where each pallet and forklift has an associated module. Pallet modules have lower power (to save battery life), and only present output in the form of small led lights (for routing). Instead, forklift modules are connected with a simple application on the workers' personal smartphone, which allows the worker to provide some basic input (i.e., logging loading and unloading details, issuing routing requests), shows her some basic output (i.e., collision warnings and additional route information), and stores locally the collected logs, uploading them on the cloud as soon as possible.

Service 1 is realised as a simple aggregate process, that is spawned by every forklift module and extends until reaching the safety radius. In this area, the distance towards the closest other forklift is gathered: if this distance decreases faster than the threshold, a warning is issued.

Service 2 is also realised through aggregate processes. One process computes routes towards empty spaces (more precisely, pallets that detect an empty space around them), others compute routes towards goods satisfying given queries, as they are issued. Each process expands naturally into the whole network, and is terminated everywhere once the routing request is cancelled.

Service 3 is realised through two simultaneous aggregate collection processes, to enable redundancy and greatly reducing the chances of information losses. Forklift modules are used as collection sinks, since they can upload data on the cloud: based on their unique identifier, half of them are assigned to sink group 1 and the other half to sink group 2. The logs produced are collected twice, towards the closest sink in group 1 and towards the closest sink in group 2. The collection algorithm used is a custom version of \emph{multi-path collection}, designed to keep the network load low for the specific log collection task. Firstly, hop-count distances towards sinks are produced. Then, every device computes its partial log collection, by including every log that appears farther than it from the sink, but \emph{not} also closer than it from the sink. This second condition ensures that logs stop being propagated when they are already closer to the sink, greatly reducing the communication load.

\subsection{Simulations}

\begin{figure}[t]
\centering
\includegraphics[width=\linewidth]{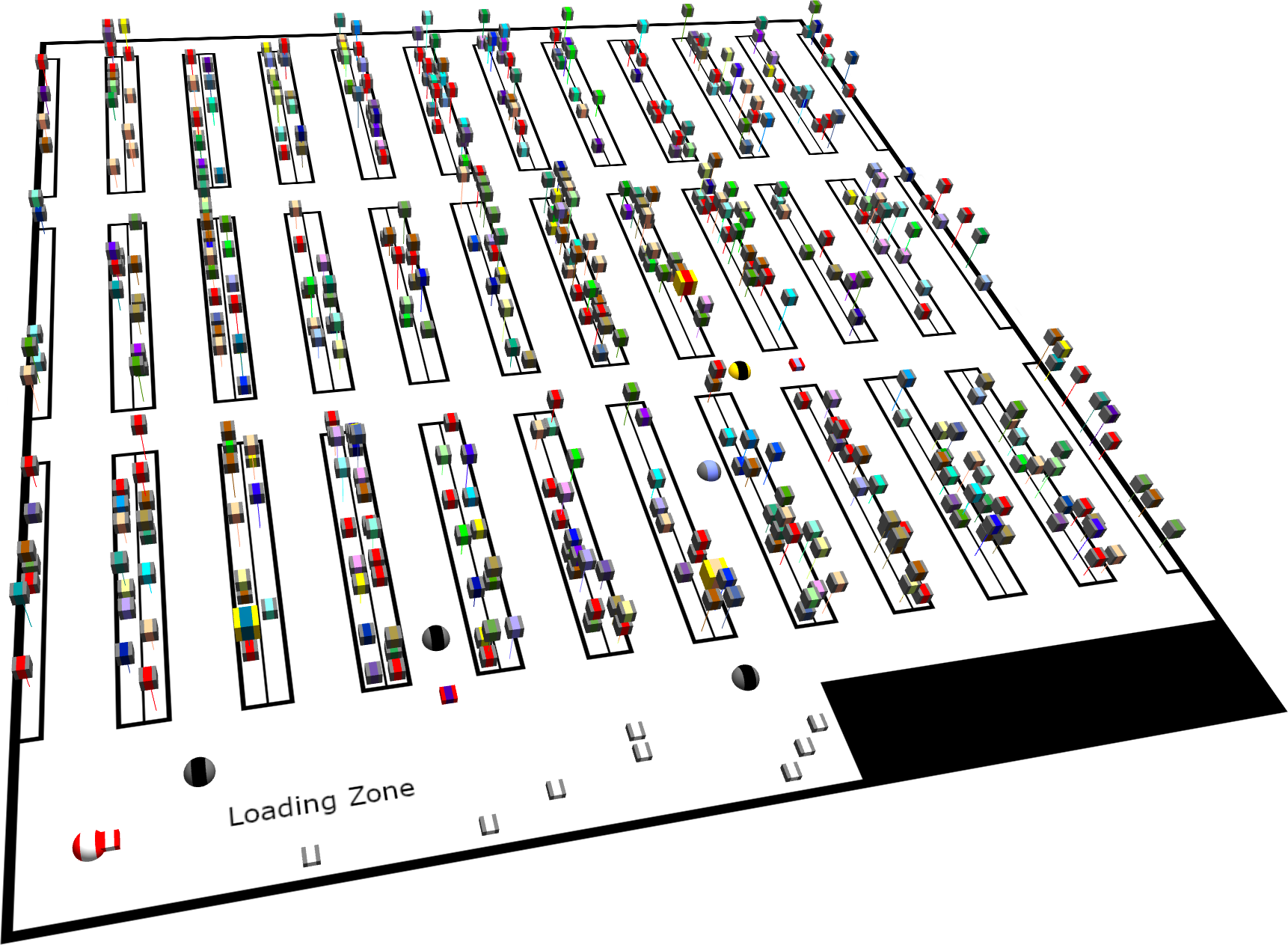}
\caption{Screenshot of the simulation execution.} \label{fig:screenshot}
\end{figure}

Firstly, we simulated the operations of a smart warehouse empowered by the proposed app through the FCPP simulation framework \cite{a:fcpp} for aggregate computing.\footnote{Code available at: \url{https://github.com/fcpp-experiments/warehouse-case-study}} A screenshot of the simulation is shown in Figure \ref{fig:screenshot}. The simulated warehouse consists of 22 rows $\times$ 3 columns of aisles; each of them composed by 15 slots in the horizontal direction $\times$ 3 slots in the vertical direction. The bottom part is dedicated to a loading zone (left) and office space (right).

Pallets are represented as cubes with a $1m$ side (spaced $1.5m$ from each other), while forklifts are represented as spheres. The content of a pallet is displayed by the colour of its middle band: white represents no content, and various colours are used for 100 different types of goods (randomly generated according to a Zipf distribution \cite{zipf}, so that few goods are very common, and many are uncommon). The middle band of a forklift is coloured analogously according to the good that the forklift is currently searching or loading (if any, black otherwise). For both pallets and forklifts, the lateral bands are grey if the device is idling, yellow if it has a led turned on (pallet routing or forklift signalling a collision risk), and red during handling (loading/unloading for forklifts, being carried by a forklift for pallets). Forklifts randomly perform either a retrieve task of a specific good (picking up a matching pallet, and bringing it to the loading zone for unload), or a insert task (where an empty palled is filled up and then brought to an empty space in the aisles). Tasks are generated randomly during idling times.

In Figure \ref{fig:screenshot}, we can see few empty pallets in the loading zone (gray and white cubes), and several hundreds of loaded pallets in the aisles (coloured cubes in the grid). In the loading zone, two forklifts are currently idling (gray and black spheres), while one is currently unloading a pallet (bottom left corner, coloured red and white).  A forklift (bottom of the 4th vertical aisle, gray and black sphere followed by a red and violet cube) has recently loaded a pallet, and is bringing it to the closest available space. Another forklift (bottom of the 7th vertical aisle, grey and cerulean sphere) is currently looking for a specific good (identified by the cerulean colour), following led lights (currently, the yellow and red cube further up in the same aisle). The last forklift (middle-bottom horizontal corridor, yellow and black sphere followed by a red and cerulean cube) is bringing back a pallet to the loading zone for unloading. Since these last two forklift are quickly approaching the same intersection, a collision warning is triggered (the external yellow bands of the last forklift).

\begin{figure}[t]
\centering
\includegraphics[width=0.49\linewidth]{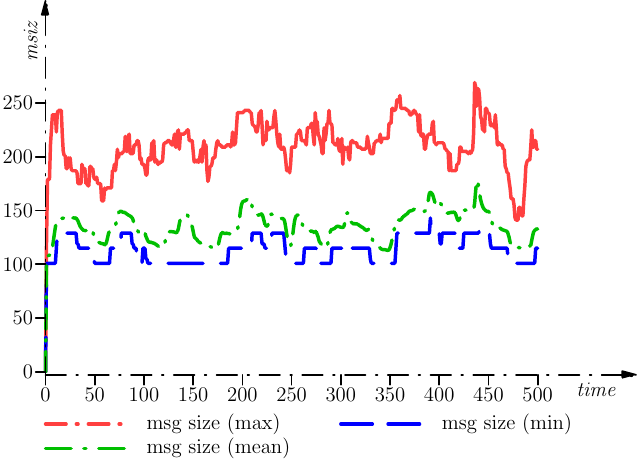}
\includegraphics[width=0.49\linewidth]{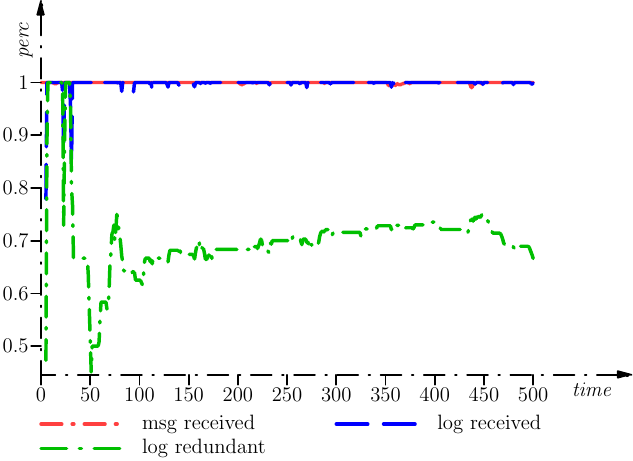} \\
\raisebox{-1.0\height}{\includegraphics[width=0.49\linewidth]{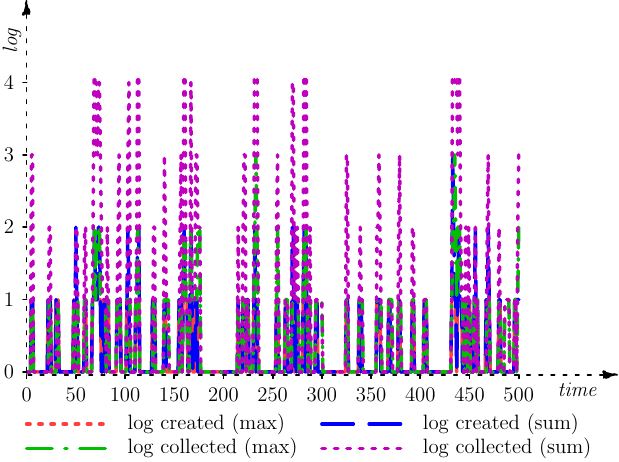}}
\raisebox{-1.0\height}{\includegraphics[width=0.49\linewidth]{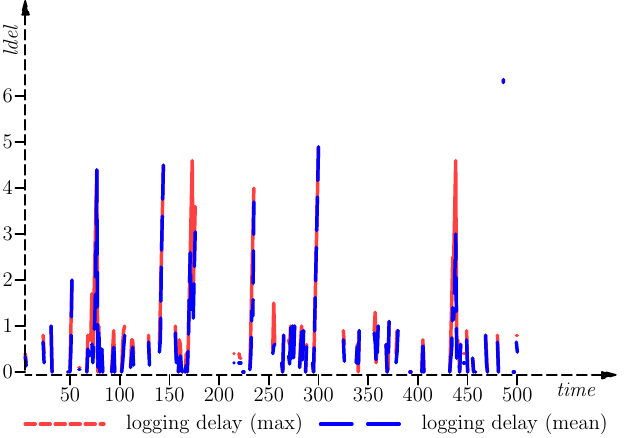}}
\caption{Plots of simulated performance over time: message size in bytes (top-left); percentage of messages delivered, log received at least once and received twice (top-right); logs created and collected (bottom-left), average delay of log collection (bottom-right).} \label{fig:plots}
\end{figure}

Figure \ref{fig:plots} presents few performance indicators of the proposed app through the course of the first 500 seconds of simulated time. The size of messages is usually below 150 bytes, with peaks below 250 bytes: almost every message is small enough to be sent, as the message limit of the modules is currently of 222 bytes (and 20 bytes are used by the simulation logics, and would not be used in a deployment). Every log that is created is eventually received, but only about 70\% of them are received twice in both sink groups, while another 30\% is only received in one sink group, proving the effectiveness of the redundancy strategy. Across the simulation, at most 2 logs are created simultaneously (and never more than one in a single device), while a single sink may collect up to 3 logs in a single round (with 4 total logs collected in the same round by sinks overall). The average delay between the log creation and collection is very small, being mostly below 5 seconds (missing parts in the bottom-right graph correspond to times when no log was collected, so that no delay was computable).

\subsection{Physical Deployment: a Proof-of-Concept}

\begin{figure}[t]
\centering
\includegraphics[width=0.4\linewidth]{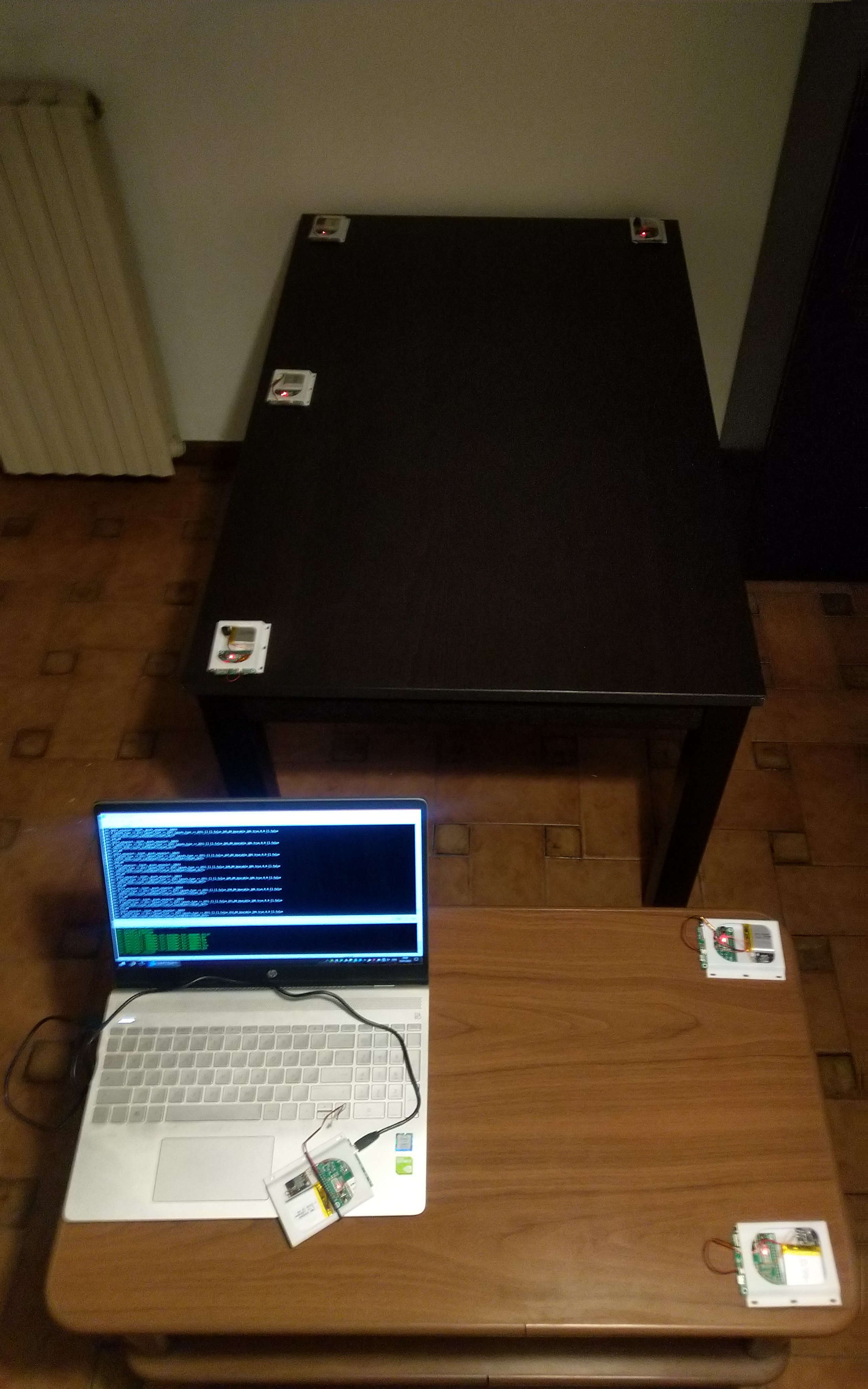}
\caption{Layout of the devices in the Proof-of-Concept} \label{fig:deployment}
\end{figure}

We validated the aggregate program implemented for the simulation on a batch of seven DWM1001-DEV modules. Six modules were configured as if they were attached to pallets, four of them having some goods already loaded and two empty, while one module was configured as the personal device of a forklift operator. The devices were arranged as visible in Figure \ref{fig:deployment}, with the modules associated to loaded pallets on the darker table, the module of the forklift operator connected to a PC to read the logs in real time and the modules associated to empty pallets near the PC. The BLE transmission power was configured to -16dBm to reduce the range of communication due to the small space available.

 Two scenarios of actions performed by the operator were tested: the loading of goods on an empty pallet followed by its placement near the already loaded pallets, and the search for a pallet containing a predermined kind of good followed by its unloading near the empty pallets. In both scenarios the operator interacts with the modules by using the button available on the personal module while receiving a feedback from the LEDs. In the former scenario the operator marks the nearest pallet as loaded with some good by using the button on the personal module, then finds an empty space near the previously loaded pallets by following the ones with the LEDs turned on. In the latter scenario the operator starts the request to find a kind of good by pressing the button on the personal module, then follows the turned on LEDs on the loaded pallets until reaching one containing the requested goods. The LEDs on that pallet start to blink and the operator delivers it back to the starting area and marks the pallet as empty by pressing the button on the pallet's module.

At the end of each test the FCPP logs, from both the personal module and the module representing the pallet handled by the operator, were collected on the PC to verify that the network presented the expected behavior. In all tests the network was left in a consistent state but all the computation were slower than expected (with reaction times of 5s to 10s) due to an high number of dropped messages in both communication and ranging.

\section{Conclusions}
\label{sec-end}

We have considered intelligent services for the \ac{iiot}, deployed on far-edge devices
  placed
  at fixed locations in the workshop floor or attached to people and moving machines, and sharing information with cloud servers through edge gateways.

Previous work  on offloading part of the computational and storage needs to the far edge of the \ac{iiot} mostly focusses on data management
and proposes architectural schemes to satisfy specific needs of efficient storage distribution or edge-to-cloud communication (c.f.\ Section \ref{sec-relwork}).
Compared to such existing work,  we show that by
exploiting the $\spawnK$ construct of FCPP
it is possible to implement any kind of service that a set of client nodes may require from a set of potential server nodes by querying the far-edge network (including safety, information retrieval, path planning). Additionally, the \ac{ap} paradigm guarantees ``for free'' that the system is open and adaptable, thus allowing node failures, mobile nodes, and nodes joining/leaving the network at any time.

A valuable contribution of the paper is the actual deployment of \ac{ap}-based applications
on physical IoT boards with highly constrained resources. Without the highly optimized FCPP library
it would have been impossible to achieve these results.
Alternative libraries implementing \ac{ap} based on the JVM \cite{cv:scafi}, would introduce a memory footprint that exceeds by far the resources of the target devices.

We want to further pursue the research described here in several directions. First of all, we would like to experiment with the physical deployment in a full-size, real world factory setting, to assess the scalability and reliability of our systems when faced with a noisy, highly dynamic environment. In particular, we envision the need for a better synchronization between FCPP computation rounds and the ranging operations, and for retransmission protocols, since with the current deployment many messages are lost. The deployment of a larger number of devices in an area corresponding to a real warehouse should also provide valuable feedback for tuning our system.

We would also like to exploit the ranging capabilities offered by the DecaWave boards to implement an efficient and accurate cooperative RTLS (Real Time Location System) based on FCPP. Such a service would open the way for many other interesting services to be offered at the far edge. In particular, we envision the possibility of designing ``intelligent'' triangulation algorithms for 3D position estimation, to be exploited in routing and discovery services.

Finally, we are considering to apply our approach to scenarios involving large numbers of mobile robotic agents that need to coordinate in an indoor space (e.g., factory, warehouse) to achieve global goals.

\section*{Acknowledgements}
The authors are grateful to Maurizio Griva and Andrés Hernando Muñoz Herrera from Reply for their advice in defining the use cases for the experiments.

\bibliography{long}

\end{document}